\documentclass[twocolumn,showkeys,showpacs,preprintnumbers,prd,superscriptaddress,nofootinbib]{revtex4-2}
\usepackage{graphicx}
\usepackage{epsf}
\usepackage{bold-extra}
\usepackage{bm}
\usepackage{amsmath}
\usepackage{amsfonts}
\usepackage{amssymb}
\usepackage{epstopdf}
\usepackage{natbib}
\usepackage{color}
\usepackage[dvipsnames,table]{xcolor}
\usepackage{physics}
\usepackage[colorlinks = true,
            linkcolor = blue,
            urlcolor  = blue,
            citecolor = blue,
            anchorcolor = blue]{hyperref}
\usepackage{verbatim}
\usepackage{lipsum}
\usepackage{float}
\usepackage{multirow}
\usepackage{soul}

\usepackage[capitalize]{cleveref}
\providecommand{\U}[1]{\protect\rule{.1in}{.1in}}

\newcommand{\be}{\begin{equation}}
\newcommand{\ee}{\end{equation}}
\newcommand{\bea}{\begin{eqnarray}}
\newcommand{\eea}{\end{eqnarray}}

\begin{document}

\title{Updated Constraints on Omnipotent Dark Energy: A Comprehensive Analysis with CMB and BAO Data}

\author{Enrico Specogna}
\email{especogna1@sheffield.ac.uk}
\affiliation{School of Mathematical and Physical Sciences, University of Sheffield, Hounsfield Road, Sheffield S3 7RH, United Kingdom}

\author{Shahnawaz A. Adil}
\email{shazadil14@gmail.com}
\affiliation{Instituto de Ciencias Físicas, UNAM, Cuernavaca 62210, México}

\author{Emre \"{O}z\"{u}lker}
\email{e.ozulker@sheffield.ac.uk}
\affiliation{School of Mathematical and Physical Sciences, University of Sheffield, Hounsfield Road, Sheffield S3 7RH, United Kingdom}

\author{Eleonora Di Valentino}
\email{e.divalentino@sheffield.ac.uk}
\affiliation{School of Mathematical and Physical Sciences, University of Sheffield, Hounsfield Road, Sheffield S3 7RH, United Kingdom}

\author{Rafael C. Nunes}
\email{rafadcnunes@gmail.com}
\affiliation{Instituto de F\'{i}sica, Universidade Federal do Rio Grande do Sul, 91501-970 Porto Alegre RS, Brazil}
\affiliation{Divis\~ao de Astrof\'isica, Instituto Nacional de Pesquisas Espaciais, Avenida dos Astronautas 1758, S\~ao Jos\'e dos Campos, 12227-010, SP, Brazil}

\author{\"{O}zg\"{u}r Akarsu}
\email{akarsuo@itu.edu.tr}
\affiliation{Department of Physics, Istanbul Technical University, Maslak 34469 Istanbul, T\"{u}rkiye}

\author{Anjan A. Sen}
\email{aasen@jmi.ac.in}
\affiliation{Centre for Theoretical Physics, Jamia Millia Islamia, New Delhi-110025, India.}

\begin{abstract}

In this work, we present updated observational constraints on the parameter space of the DMS20 dark energy model, a member of the \textit{omnipotent dark energy} (ODE) class. Our analysis combines multiple CMB datasets—including measurements from the Planck satellite (PL18), the South Pole Telescope (SPT), and the Wilkinson Microwave Anisotropy Probe (WMAP)—with Type Ia supernova data from the Pantheon$+$ catalog (PP), and baryon acoustic oscillation (BAO) measurements from the DESI and SDSS surveys. We find that certain data combinations, such as SPT+WMAP+BAO and PL18+BAO, can reduce the significance of the $H_0$ tension below $1\sigma$, but with considerably large uncertainties. However, the inclusion of PP data restores the tension in $H_0$. To provide a comprehensive view of the ODE phenomenology, we also investigate the evolution of its energy density, emphasizing its dynamical behavior at low redshifts. Our results generically exhibit multiple phantom divide line crossings in a single expansion history; if confirmed, this points beyond the simplest minimally coupled canonical single-field quintessence/phantom descriptions and motivates more general dark-sector realizations.

\end{abstract}

\maketitle

\section{Introduction}

The $\Lambda$CDM model has emerged as the dominant and most widely accepted cosmological framework. Its success is largely attributed to its remarkable accuracy in explaining a broad range of astrophysical and cosmological observations without excessive model complexity. Despite these successes, however, the $\Lambda$CDM model faces some unresolved challenges, particularly in capturing a consistent expansion history when various combinations of mainstream cosmological observables, such as the cosmic microwave background (CMB), baryon acoustic oscillations (BAO), large-scale structure (LSS), and local distance ladder measurements of the Hubble constant are simultaneously considered~\cite{Perivolaropoulos:2021jda,Abdalla:2022yfr,DiValentino:2022fjm,Peebles:2024txt,DiValentino:2025sru}.

With the increasing precision of modern observations~\cite{DiValentino:2020vhf}, it is anticipated that deviations from the standard $\Lambda$CDM model will become more apparent. Indeed, various discrepancies in the estimation of key cosmological parameters have already emerged, some of which show statistically significant departures from the predictions of the model~\cite{Abdalla:2022yfr,DiValentino:2020vvd,DiValentino:2020srs,DiValentino:2020zio,Perivolaropoulos:2021jda,Akarsu:2024qiq,DiValentino:2025sru}. Among these issues, one of the most prominent and statistically significant is the ongoing discrepancy related to the Hubble constant, $H_0$ (for a detailed review, see~\cite{Verde:2019ivm,DiValentino:2020zio,DiValentino:2021izs,Perivolaropoulos:2021jda,Shah:2021onj,Abdalla:2022yfr,DiValentino:2022fjm,Kamionkowski:2022pkx,Verde:2023lmm,DiValentino:2024yew,Perivolaropoulos:2024yxv,DiValentino:2025sru} and references therein). A notable conflict exists between the values of $H_0$ inferred from the CMB measurements~\cite{Planck:2018vyg,SPT-3G:2022hvq,ACT:2025fju} assuming the $\Lambda$CDM model, and those derived from direct, independent measurements obtained through local astrophysical observations~\cite{Freedman:2020dne,Birrer:2020tax,Wu:2021jyk,Anderson:2023aga,Scolnic:2023mrv,Jones:2022mvo,Anand:2021sum,Freedman:2021ahq,Uddin:2023iob,Huang:2023frr,Pesce:2020xfe,Kourkchi:2020iyz,Schombert:2020pxm,Blakeslee:2021rqi,deJaeger:2022lit,Murakami:2023xuy,Breuval:2024lsv,Freedman:2024eph,Riess:2024vfa,Vogl:2024bum,Gao:2024kkx,Scolnic:2024hbh,Said:2024pwm,Boubel:2024cqw,Scolnic:2024oth,Li:2025ife,Jensen:2025aai}. This tension, commonly referred to as the $H_0$ tension, has reached a statistical significance greater than $5\sigma$~\cite{Riess:2021jrx,Murakami:2023xuy,Breuval:2024lsv}, and is now considered one of the most pressing issues within the context of the $\Lambda$CDM model. To solve this tension, cosmologists have been exploring both possible systematic errors~\cite{Efstathiou:2020wxn,Mortsell:2021nzg,Mortsell:2021tcx,Riess:2021jrx,Sharon:2023ioz,Murakami:2023xuy,Riess:2023bfx,Bhardwaj:2023mau,Brout:2023wol,Dwomoh:2023bro,Uddin:2023iob,Riess:2024ohe,Freedman:2024eph,Riess:2024vfa} and alternative cosmological models~\cite{Knox:2019rjx,Jedamzik:2020zmd,DiValentino:2021izs,Perivolaropoulos:2021jda,Schoneberg:2021qvd,Abdalla:2022yfr,Kamionkowski:2022pkx,Khalife:2023qbu,Giare:2023xoc,Hu:2023jqc,DiValentino:2024yew,DiValentino:2025sru}. In addition to the $H_0$ tension, another point of discussion in cosmology is the $S_8$ tension~\cite{Nunes:2021ipq,Abdalla:2022yfr,DiValentino:2020vvd,DiValentino:2018gcu} (see also~\cite{Adil:2023jtu,Akarsu:2024hsu}), which concerns discrepancies between the amplitude of matter fluctuations inferred from early-universe measurements, such as the Planck CMB data, and the $S_8$ values derived from late-time observations, including weak gravitational lensing and galaxy clustering, though it seems to have lost significance after the new KiDS-Legacy release~\cite{DES:2021bvc,DES:2021vln,KiDS:2020suj,Asgari:2019fkq,Joudaki:2019pmv,DAmico:2019fhj,Kilo-DegreeSurvey:2023gfr,Troster:2019ean,Heymans:2020gsg,Dalal:2023olq,Chen:2024vvk,ACT:2024okh,Sailer:2024coh,Harnois-Deraps:2024ucb,Dvornik:2022xap,DES:2021wwk,Wright:2025xka}.
Furthermore, recent results from the DESI collaboration analyzing the BAO feature have found evidence for dynamical dark energy (DDE) when their measurements are combined with other major observations~\cite{DESI:2024mwx,DESI:2024aqx,DESI:2024jis,DESI:2025zgx,DESI:2025fii}. This evidence for DDE is relevant only in the post-recombination era, with a statistical significance of $3.1\sigma$ when combined with the CMB. The significance ranges from $2.7\sigma$ to $4.2\sigma$ when different supernovae samples are included in the dataset combination. The indication persists even without the CMB; for example, only the DES supernovae sample~\cite{DES:2024jxu} when combined with DESI BAO yields a $3.3\sigma$ significance. It is important to note that the evidence for DDE does not hinge on DESI BAO either, e.g., see Refs.~\cite{DES:2024jxu,DES:2025bxy} that report preference of a DDE without this data set. A plethora of models have been proposed in the literature to address the existing cosmological tensions~\cite{DiValentino:2025sru}, a substantial portion of which incorporate a DDE component that is non-negligible in the late and/or early universe~\cite{DESI:2024mwx,Cortes:2024lgw,Shlivko:2024llw,Luongo:2024fww,Yin:2024hba,Gialamas:2024lyw,Dinda:2024kjf,Najafi:2024qzm,Wang:2024dka,Tada:2024znt,Carloni:2024zpl,Chan-GyungPark:2024mlx,DESI:2024kob, Bhattacharya:2024hep,Ramadan:2024kmn,
Bhattacharya:2024kxp,Orchard:2024bve,Hernandez-Almada:2024ost,Pourojaghi:2024tmw,Giare:2024gpk,Reboucas:2024smm,Giare:2024ocw,Chan-GyungPark:2024brx,Menci:2024hop,Li:2024qus,Li:2024hrv,Notari:2024zmi,Gao:2024ily,Fikri:2024klc,Jiang:2024xnu,Zheng:2024qzi,Gomez-Valent:2024ejh,RoyChoudhury:2024wri,RoyChoudhury:2025dhe,Lewis:2024cqj,Wolf:2025jlc,Shajib:2025tpd,Giare:2025pzu,Chaussidon:2025npr,Kessler:2025kju,Alvarez:2025kma}.

The simplest phenomenological DDE model replaces the cosmological constant, whose effective equation of state (EoS) parameter is $w = -1$, with a DE fluid that has a constant EoS parameter able to satisfy $w_{\rm DE}\neq-1$; this is often referred to as the $w$CDM model. Many models extend this approach by allowing a varying EoS parameter, with the most well-known example being the Chevallier-Polarski-Linder (CPL) parametrization~\cite{Chevallier:2000qy, Linder:2002et}, in which $w_{\rm DE}$ is assumed to be a linear function of the scale factor in the Robertson–Walker (RW) metric. Notably, unless restricted to specific regions of parameter space, such extensions generically allow $w_{\rm DE}$ to cross the phantom divide line (PDL), $w_{\rm DE} = -1$. However, parameterizing DDE solely via its EoS parameter fails to capture the phenomenology of models in which DDE attains negative effective energy densities in the past, as such components would exhibit a singular EoS parameter~\cite{Ozulker:2022slu}. From both theoretical and phenomenological perspectives, models predicting negative energy densities in the dark sector have garnered increasing attention in recent years, particularly for their potential to resolve the $H_0$ tension~\cite{Dutta:2018vmq,Visinelli:2019qqu,Malekjani:2023ple,Vazquez:2023kyx,Adil:2023ara,Gomez-Valent:2023uof,Akarsu:2021fol,Akarsu:2022typ,Akarsu:2024qsi,Akarsu:2024eoo,Akarsu:2019hmw,Akarsu:2023mfb,Yadav:2024duq,Anchordoqui:2023woo,Menci:2024rbq,Anchordoqui:2024gfa,Anchordoqui:2024dqc,Soriano:2025gxd,DeFelice:2023bwq,Gomez-Valent:2024tdb,DESI:2024aqx,Escamilla:2024ahl,Escamilla:2023shf,Sabogal:2024qxs,Escamilla:2024ahl,Akarsu:2024nas,Akarsu:2025gwi,DiValentino:2025sru}. DDE models capable of simultaneously incorporating all of these features were dubbed \textit{omnipotent dark energy} (ODE) models in Ref.~\cite{Adil:2023exv}, where it was argued that such a flexible DDE component may be necessary for a satisfactory resolution of the prevailing cosmological tensions.

In Ref.~\cite{DiValentino:2020naf}, the DMS20 model was proposed as a promising solution to the cosmological tensions and was found to prefer a PDL crossing at $z \sim 0.1$. The recent analysis in Ref.~\cite{Adil:2023exv} identified DMS20 as an ODE model and showed that its ability to reach negative energy densities for $z \gtrsim 2$—mimicking a negative cosmological constant at high redshifts—plays a crucial role in alleviating the tensions. This behavior is consistent with the predictions of the $\Lambda_{\rm s}$CDM model~\cite{Akarsu:2021fol, Akarsu:2022typ, Akarsu:2023mfb}, which suggests a transition from an Anti-de Sitter (AdS) to a de Sitter (dS) phase. This transition can be interpreted either as an emergent effect from modified gravity or as the result of an actual field within the framework of general relativity (GR).

In this paper, we revisit and update the observational constraints on the DMS20 model using Type Ia supernova data from the Pantheon$+$ catalog, BAO measurements from the DESI and SDSS surveys, and CMB temperature and polarization data from Planck, the South Pole Telescope, and the Wilkinson Microwave Anisotropy Probe. Motivated by previous studies of the DMS20 model that reported posterior distributions leaning against the edges of the prior ranges for certain dataset combinations, we extend these prior ranges. This not only relaxes the constraints on the model parameters but also enables the model to exhibit richer phenomenology that is qualitatively different—for example, a crossing of the PDL, occurring either from the quintessence regime to the phantom regime or vice versa. Nevertheless, we also present our results after excluding the MCMC samples corresponding to the extended priors, to facilitate direct comparison with previous studies.

The structure of this paper is organized as follows. In~\cref{model}, we review the key physical features of the omnipotent dark energy model, focusing on the aspects most relevant for cosmological tests. In~\cref{data}, we describe the observational datasets used in our analysis and outline the methodology employed to constrain the model’s free parameters. In~\cref{results}, we present and discuss our main results, highlighting their implications for cosmological tensions and the broader landscape of dark energy models. Finally, in~\cref{final}, we summarize our conclusions and offer some perspectives on future research directions.

\section{Omnipotent Dark Energy: a review}
\label{model}
In this section, we review the class of ODE models and the DMS20 model as a concrete member of this class. The term ODE describes a family of phenomenological proposals that allow energy densities to transition between positive and negative values, and to exhibit oscillatory or non-monotonic evolution histories. ODE models permit arbitrary EoS, including singularities and PDL crossings, unconstrained by standard energy conditions. As an effective source in the Friedmann equations, ODE provides a flexible framework to address the limitations of the cosmological constant. More precisely, a DE model is referred to as an ODE if, for any point in its parameter space, it can exhibit all six possible combinations of the conditions $\rho_{\rm DE} > 0$, $\rho_{\rm DE} < 0$, and $w_{\rm DE} > -1$, $w_{\rm DE} = -1$, $w_{\rm DE} < -1$ within a single expansion history. Here, $\rho_{\rm DE}$ denotes the energy density of the DE component~\cite{Adil:2023exv}.

Two distinctive features of ODE are its capacity to realize a non-monotonic energy density evolution and to reach negative energy densities, as suggested by various observational reconstructions~\cite{Bonilla:2020wbn,Sahni:2014ooa,Aubourg:2014yra,Wang:2018fng,Poulin:2018zxs,Escamilla:2021uoj,Escamilla:2023shf}. Such behavior has been shown to alleviate key cosmological tensions, including the $H_0$ and $S_8$ discrepancies~\cite{Visinelli:2019qqu,Sen:2021wld,Calderon:2020hoc,Sahni:2014ooa,DiValentino:2020naf,Akarsu:2019hmw,Dutta:2018vmq,Akarsu:2021fol,Akarsu:2022typ,Akarsu:2019ygx,Acquaviva:2021jov,Akarsu:2022lhx,Vazquez:2023kyx}.
Hereafter, we consider an ODE component, whether as an effective source from modified gravity or a fluid within GR, that satisfies the usual continuity equation that follows from the local conservation of the energy-momentum tensor for the RW metric. Given this weak assumption, unlike usual phantom DE~\cite{Caldwell:1999ew}, which maintains $w_{\rm DE} < -1$, the non-monotonicity feature of ODE models requires that its EoS parameter crosses the PDL. Moreover, since its energy density vanishes during the transition between negative and positive density regions at a scale factor $a_{\rm p}$, its EoS parameter exhibits a pole of the form $\lim_{a \to a_{\rm p}^\pm} w_{\rm DE}(a) = \mp \infty$; this also corresponds to a PDL crossing, but a discontinuous one~\cite{Ozulker:2022slu}.

The DMS20 model proposed in Ref.~\cite{DiValentino:2020naf} serves as a concrete example of an ODE model. In this model, the DE density, $\rho_{\rm DE}$, is parameterized to ensure an extremum at a scale factor $a_m$, satisfying the condition $\left. \frac{{\rm d} \rho_{\rm DE}}{{\rm d}a} \right|_{a = a_m} = 0$. The DE density is then expressed by expanding $\rho_{\rm DE}$ around $a_m$:
\begin{equation}
    \rho_{\rm DE}(a) = \rho_{\rm DE0} \frac{1 + \alpha (a - a_m)^2 + \beta (a - a_m)^3}{1 + \alpha (1 - a_m)^2 + \beta (1 - a_m)^3},
    \label{eq:rho_ode}
\end{equation}
where $\alpha$ and $\beta$ are constants defining the polynomial terms (for further details, see Refs.~\cite{DiValentino:2020naf,Adil:2023exv}). The absence of a linear term in the expansion follows from the vanishing of the first derivative at $a = a_m$. The parameter $a_m$ has significant physical implications: from the continuity equation, it follows that $w_{\rm DE}(a_m) = -1$, provided that $\rho_{\rm DE}$ is non-zero at the extremum. For $\alpha > 0$, this corresponds to a transition from $w_{\rm DE} > -1$ to $w_{\rm DE} < -1$ as the universe expands, while for $\alpha < 0$, the crossing occurs in the opposite direction. The EoS for the ODE model is given by:
\begin{equation}
\label{EoS_model}
w_{\rm DE}(a) = -1 - \frac{a[2\alpha(a - a_m) + 3\beta(a - a_m)^2]}{3[1 + \alpha(a - a_m)^2 + \beta(a - a_m)^3]},
\end{equation}
which yields $w_{\rm DE}(a = 0) = -1$ and $w_{\rm DE}(a \to \infty) = -2$.

\begin{figure*}
    \centering
    \includegraphics[width=1.\linewidth, trim=105 120 120 60, clip]{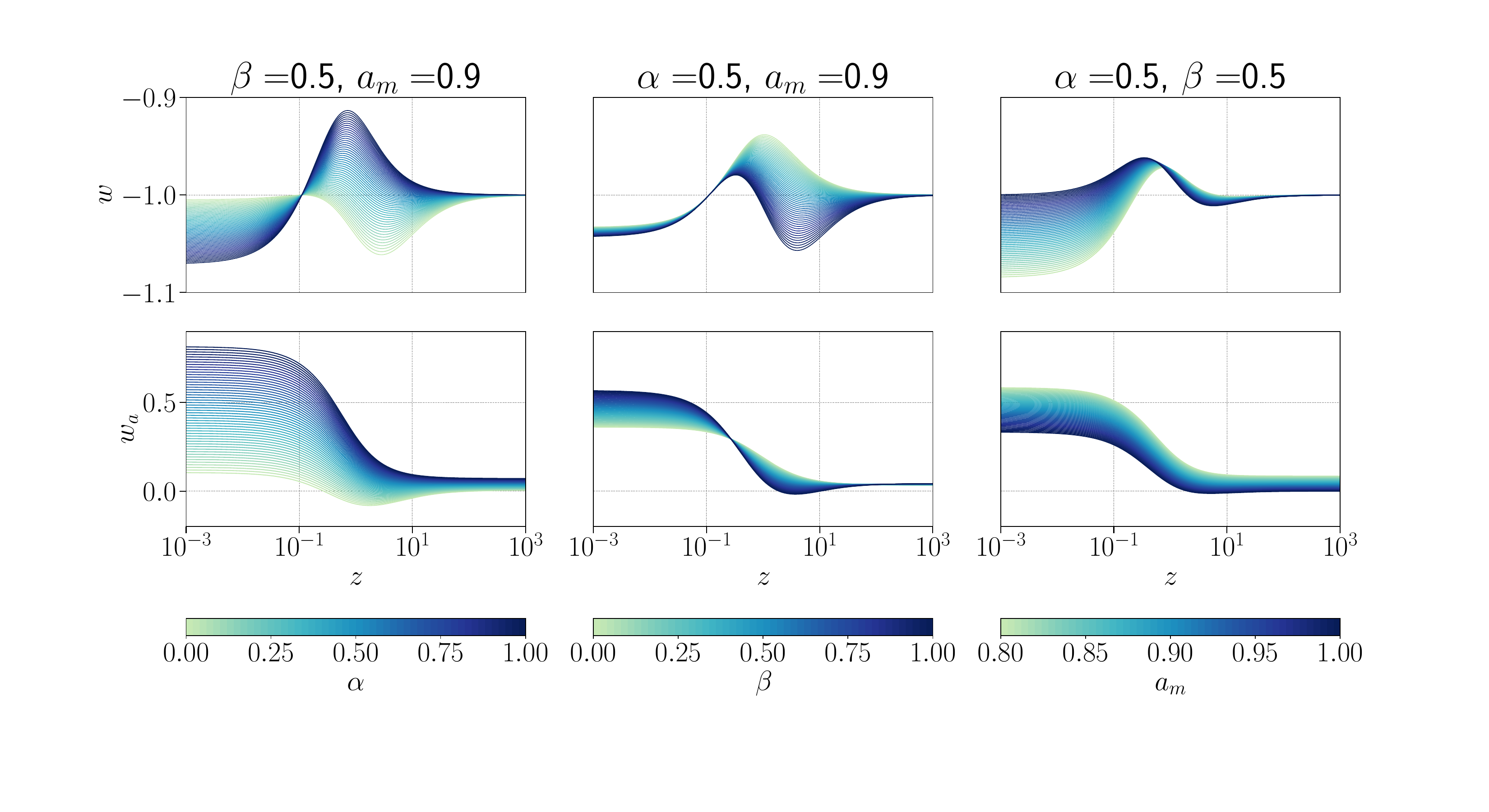}
    \caption{The redshift evolution of $w_{\rm DE}$ and the CPL parameter $w_a$ are shown for different values of $\{a_m, \alpha, \beta\}$. Clearly $w_a$ cannot remain constant in this model, so the linear CPL parametrization of $w_{\rm DE}$ fails to capture the oscillatory behavior characteristic of ODE.}
    \label{w0wa}
\end{figure*}

This model introduces three extra parameters: $\{a_m, \alpha, \beta\}$. Depending on the values of these parameters, certain features of ODE may remain dormant. For instance, when $\alpha = \beta = 0$, the DE density reduces to the cosmological constant $\Lambda$, recovering the standard $\Lambda$CDM model. For a detailed discussion of the dynamical behavior of the model across different values of $\{a_m, \alpha, \beta\}$, see Ref.~\cite{Adil:2023exv}. Here, we simply note that the EoS of DMS20 cannot, in general, be represented \emph{exactly} by the CPL parametrization, $w_{\rm DE} = w_0 + w_a z / (1 + z)$, where $z = (1 - a)/a$ is the redshift. Here, $w_0 = w_{\rm DE}(z=0)$, and $w_a$ controls the leading time-variation of $w_{\rm DE}$ within the CPL ansatz (equivalently, $w(a)=w_0+w_a(1-a)$, so ${\rm d}w/{\rm d}a=-w_a$ is constant by construction). More generally, $(w_0,w_a)$ are best regarded as a convenient two-parameter \emph{data-compression} of the model predictions for a given set of observables and redshift sensitivities (see, e.g., \cite{Garcia-Garcia:2019cvr,Wolf:2023uno,Wolf:2024eph}). These two parameters are essential for characterizing the behavior of dark energy across a wide range of models, as discussed in~\cite{linder2024}. However, as shown in~\cref{w0wa}, for any choice of $\{a_m, \alpha, \beta\}$ we observe an oscillatory pattern in $w_{\rm DE}$, which cannot be described by the linear CPL form. In fact, $w_a$ would need to acquire a redshift dependence to account for the behavior of the EoS in~\cref{EoS_model}. It is seen from~\cref{w0wa} that variations in $\alpha$ significantly influence the present-day values of both $w$ and $w_a$, while having a negligible effect at higher redshifts. Specifically, as $\alpha$ increases, $w$ shifts further into the phantom regime, and $w_a$ also increases. In the middle panel, we observe that $\beta$ induces more pronounced oscillations in $w$, although its effect on $w_a$ is relatively minor. Lastly, the parameter $a_m$ shows a similar trend to $\alpha$, highlighting its role in shaping the present-day dynamics of dark energy.  It is evident that the linear CPL parametrization fails to capture the non-linear features of the ODE model, whose oscillatory behavior cannot be reproduced by such a simplified form. With only one additional parameter, our ODE parametrization is thus able to recover a complementary class of DE models with a richer phenomenology. 

To understand the background evolution of this model, we begin with the expansion rate of the universe, governed by the modified Friedmann equation:
\begin{equation}
\label{H_z_model}
\frac{H^2(a)}{H_0^2} = \Omega_{\rm m0} a^{-3} + \Omega_{\rm r0} a^{-4} + \Omega_{\rm DE0} f(a),
\end{equation}
where
\begin{equation}
f(a) = \frac{1 + \alpha (a - a_m)^2 + \beta (a - a_m)^3}{1 + \alpha (1 - a_m)^2 + \beta (1 - a_m)^3}.
\end{equation}
In Eq.~(\ref{H_z_model}), the subscript ``$m$'' refers to all forms of matter (including both baryonic and cold dark matter), while ``$r$'' denotes radiation (photons and other relativistic relics). The density parameters $\Omega_{i0} \equiv \rho_{i0} / (3 H_0^2)$ represent the present-day values of the respective energy densities.

The DMS20 model does not introduce modifications to other sectors of the universe or its constituent species. Consequently, the background evolution remains unchanged for all components other than DE itself. The linear evolution of DE perturbations follows the standard prescription, where we impose synchronous gauge conditions for metric perturbations. The continuity and Euler equations governing the DE fluid are given by
\bea
\label{laprimsim}
\dot{\delta}_x &=& -(1+w) \left(\theta + \frac{\dot{h}}{2} \right) 
- 3(\hat{c}_s^2 - w) \mathcal{H} \delta_x  \nonumber \\ 
&& - 9(1+w)(\hat{c}_s^2 - c_a^2) \mathcal{H}^2 \frac{\theta_x}{k^2}, \\  
\label{fortheta}
\dot{\theta}_x &=& -(1 - 3\hat{c}_s^2) \mathcal{H} \theta_x 
+ \frac{\hat{c}_s^2 k^2}{1 + w} \delta_x - k^2 \sigma_x.
\eea
These equations are quite general, as they assume only a non-interacting fluid and allow for the presence of shear stress $\sigma_x$, a non-adiabatic sound speed, and a time-dependent equation of state parameter $w$. Henceforth, we assume the fluid is shear-free and that $w$ is given by Eq.~(\ref{EoS_model}).

\section{Data and Methodology}
\label{data}


\begin{table}[h]
    \centering
    \caption{Flat priors are adopted to test the ODE model against all of the likelihood combinations presented in Sec.~\ref{results}, with the exception of those involving the \texttt{SPT+WMAP} dataset combination, where a Gaussian prior is imposed on the optical depth: $\mathcal{P}(\tau) = \mathcal{N}(0.0544,\ 0.0073^2)$.}
    \renewcommand{\arraystretch}{1.2}
    \resizebox{0.5\columnwidth}{!}{%
    \begin{tabular}{lc}
        \hline\hline
        Parameter & Prior \\
        \hline
        $\Omega_{\rm b} h^2$       & $[0.017,\ 0.027]$ \\
        $\Omega_{\rm c} h^2$       & $[0.09,\ 0.15]$ \\
        $\tau$                     & $[0.01,\ 0.8]$ \\
        $n_s$                      & $[0.9,\ 1.1]$ \\
        $\log(10^{10} A_s)$        & $[2.6,\ 3.5]$ \\
        $100\,\theta_{\rm MC}$     & $[1.03,\ 1.05]$ \\
        $\alpha$                   & $[-8,\ 8]$ \\
        $\beta$                    & $[-8,\ 8]$ \\
        $a_m$                      & $[0,\ 1.4]$ \\
        \hline
    \end{tabular}
    }
    \label{tab:priors}
\end{table}


We generate theoretical predictions for the ODE model using a modified version of the Boltzmann solver \texttt{CAMB}~\cite{Lewis:1999bs, Howlett:2012mh}, while parameter estimation is performed with the publicly available sampler \texttt{Cobaya}~\cite{Torrado:2020dgo}. The sampling of the posterior distributions is carried out using the Monte Carlo Markov Chain (MCMC) method, originally developed for \texttt{CosmoMC}~\cite{Lewis:2002ah}. This implementation incorporates the “fast dragging” technique~\cite{neal2005takingbiggermetropolissteps, Lewis:2013hha}, which improves efficiency in exploring parameter spaces with varying computational complexity.

In varying combinations, the likelihoods employed in this analysis use data from the following cosmological surveys:

\begin{itemize}

\item \textbf{Planck 2018 (\texttt{PL18}):} We include the full Planck 2018 temperature and polarization likelihoods (TT, TE, EE), along with the Planck 2018 lensing likelihood~\cite{Planck:2018vyg}, reconstructed from the four-point correlation function of temperature fluctuations.

\item \textbf{South Pole Telescope (\texttt{SPT}):} We incorporate CMB temperature and polarization anisotropy measurements (TT, TE, EE) from the SPT collaboration~\cite{SPT-3G:2021eoc,SPT-3G:2022hvq}.

\item \textbf{Wilkinson Microwave Anisotropy Probe (\texttt{WMAP}):} We utilize the CMB temperature and polarization data from the WMAP 9-year release~\cite{Hinshaw_2013}. 
To mitigate dust contamination, we exclude WMAP low-$\ell$ polarization data by setting the minimum multipole for TE and EE to $\ell=24$. For this reason, whenever we use the WMAP spectra we consistently apply a Gaussian prior on the optical depth to reionization, $\tau = 0.0544 \pm 0.0073$, which is the baseline choice used by the SPT--3G collaboration when they include the WMAP data in their analysis ~\cite{SPT-3G:2021eoc}. We also assume that there is no correlation between the SPT and WMAP datasets, as they probe largely different regions of the sky~\cite{SPT-3G:2021eoc}.

\item \textbf{Dark Energy Spectroscopic Instrument (\texttt{DESI}):} We include baryon acoustic oscillation (BAO) measurements from the DESI collaboration, based on galaxy and quasar observations~\cite{desicollaboration2024desi2024iiibaryon}, as well as Lyman-$\alpha$ tracers~\cite{desicollaboration2024desi2024ivbaryon}, compiled in Table I of Ref.~\cite{DESI:2024mwx}. These measurements span the redshift range \texttt{$0.1 < z < 4.2$}, divided into seven bins, and include both isotropic and anisotropic BAO constraints. The isotropic BAO measurements are expressed as $D_V(z)/r_d$, where $D_V$ is the volume-averaged distance normalized to the comoving sound horizon at the drag epoch, $r_d$. The anisotropic constraints include $D_M(z)/r_d$ and $D_H(z)/r_d$, where $D_M$ is the comoving angular diameter distance and $D_H$ the Hubble horizon. Correlations between $D_M/r_d$ and $D_V/r_d$ are accounted for.

\item \textbf{Sloan Digital Sky Survey (\texttt{eBOSS}):} We incorporate the galaxy, quasar and Lyman-$\alpha$ measurements of $D_M/r_d$ and $D_H/r_d$ from the eBOSS experiment, as presented in Table 3 of Ref.~\cite{Alam_2021}.

\item \textbf{Pantheon+ (\texttt{PP}):} We include distance modulus measurements of Type Ia supernovae from the Pantheon+ sample~\cite{Brout:2022vxf}, consisting of 1,550 supernovae over the redshift range \texttt{$0.001 \leq z \leq 2.26$}. 

\end{itemize}

For the DMS20 model parameters, we adopt the agnostic, flat priors outlined in Table~\ref{tab:priors}, except, as noted above, for an informative Gaussian prior on $\tau$ whenever the \texttt{SPT+WMAP} dataset combination is considered. Specifically, the parameter space of the DMS20 model extends the standard $\Lambda$CDM framework by introducing the three parameters defined in Eq.~\eqref{eq:rho_ode}, namely $\alpha$, $\beta$, and $a_m$, in addition to the six baseline $\Lambda$CDM parameters: the physical baryon and cold dark matter densities ($\Omega_{\rm b} h^2$, $\Omega_{\rm c} h^2$), the optical depth to reionization ($\tau$), the amplitude and spectral index of the primordial scalar fluctuations ($\log(10^{10} A_s)$, $n_s$), and the angular size of the sound horizon at last scattering ($\theta_{\rm MC}$). 
Compared to Refs.~\cite{Adil:2023exv,DiValentino:2020naf}, the priors on $\alpha$ and $\beta$ are extended to include negative values, and $a_m$ is allowed to exceed $1$, thereby enabling the possibility of PDL crossings occurring in the future (i.e., for $a > 1$). These choices are discussed in more detail in the next section.

All of the 1D and 2D posteriors were calculated and visualized using the \texttt{Getdist} code~\cite{Lewis:2019xzd}, while the functional posteriors for $w_{\rm DE}(z)$ and $\rho_{\rm DE}(z)$ shown in the next section were produced using the \texttt{fgivenx} plotting package~\cite{fgivenx}.

\begin{figure*}[t]
   \includegraphics[width=0.9\textwidth]{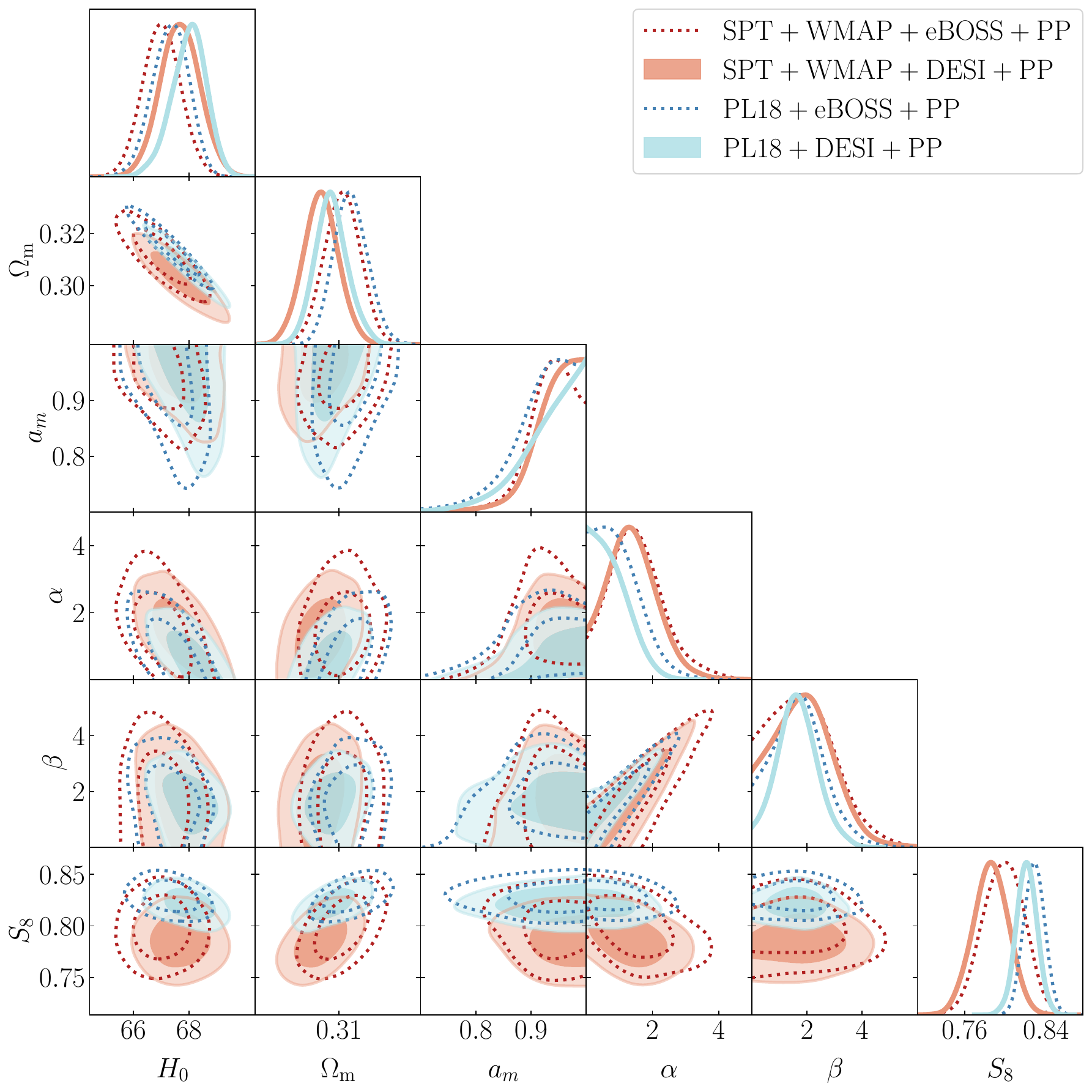}
     \caption{1D and 2D marginalized constraints on the ODE parameter space, derived from the post-filtered chains that include only samples with $\alpha > 0$, $\beta > 0$, and $a_m \in [0,1]$.}
    \label{fig:cut}
\end{figure*}

\begin{figure*}[t]
   \includegraphics[width=0.9\textwidth]{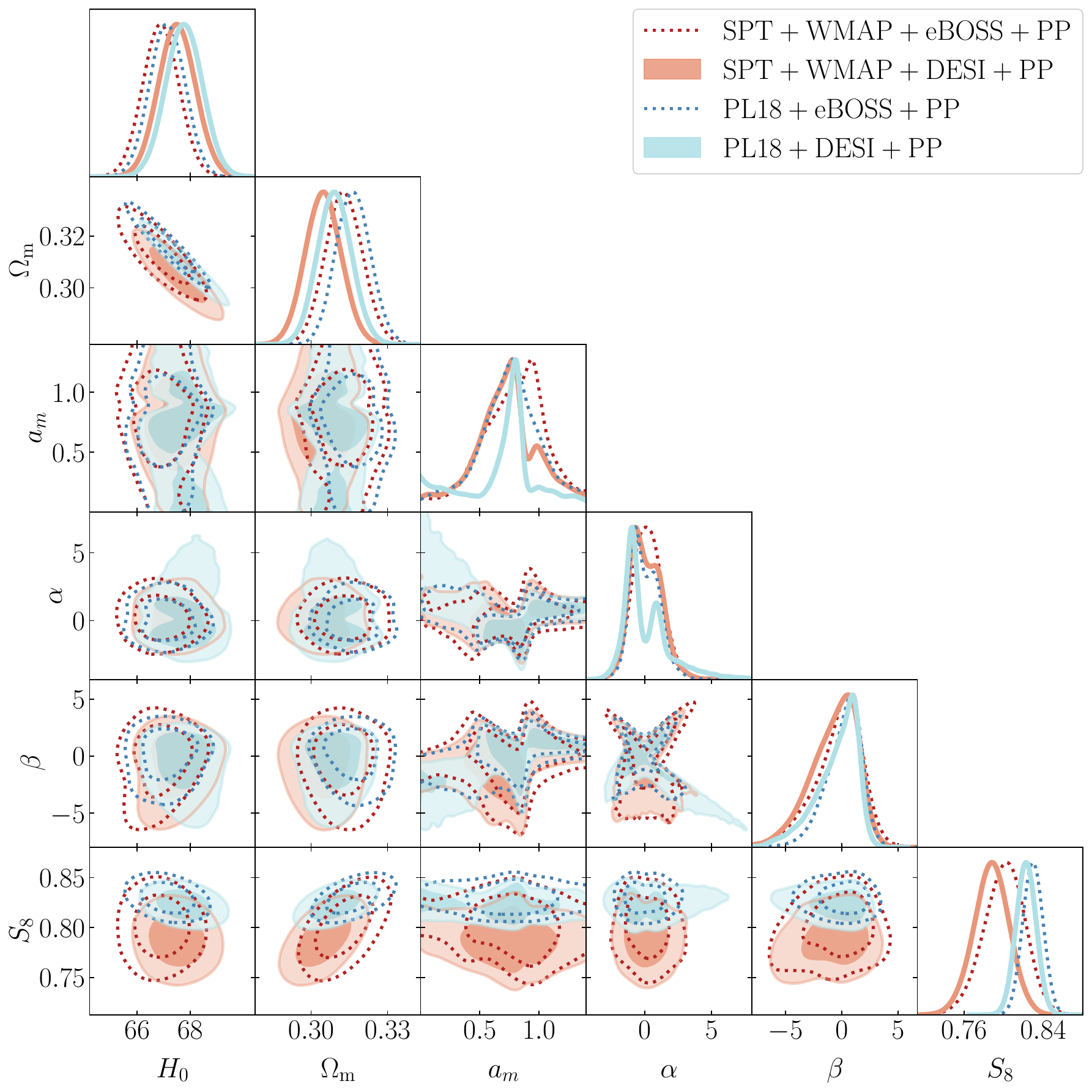}
     \caption{1D and 2D marginalized constraints on the ODE parameter space, obtained from the chains sampled with the full-prior ranges listed in Table~\ref{tab:priors}, without any post-processing cuts (full-prior results).}
    \label{fig:full}
\end{figure*}


\begin{figure*}[t]
\centering
    \includegraphics[width=1.\linewidth, trim=105 0 120 50, clip]
   {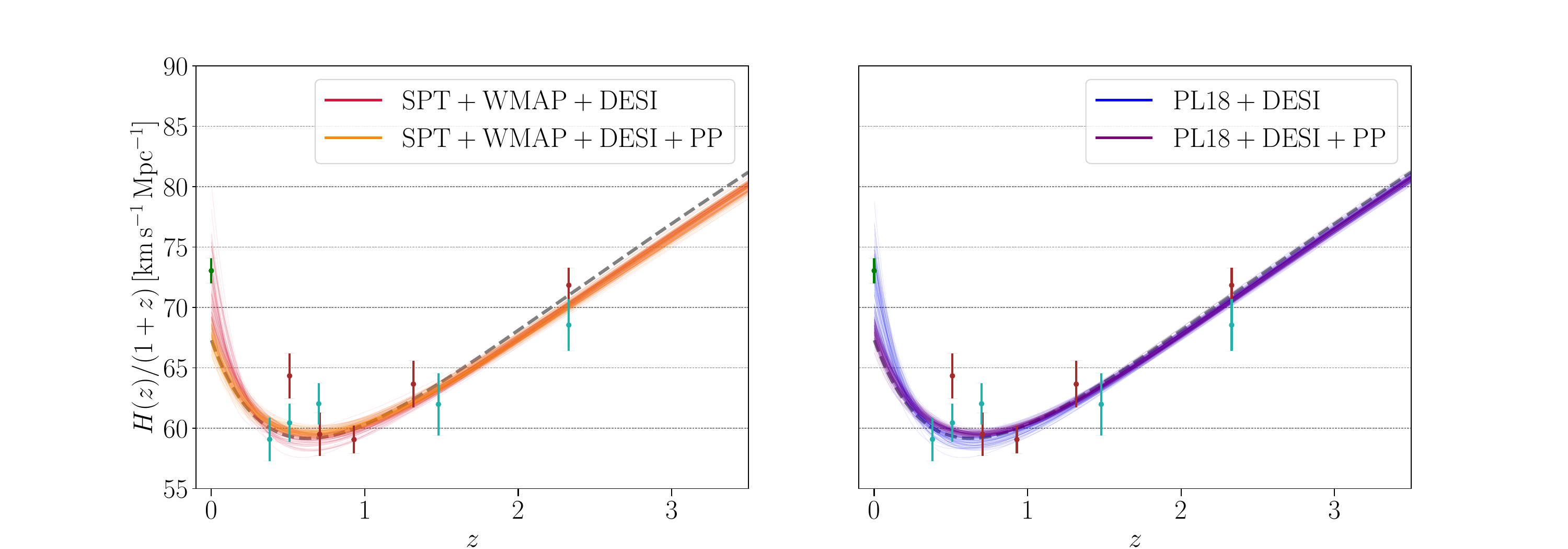}
   \includegraphics[width=1.\linewidth, trim=105 0 120 50, clip]
   {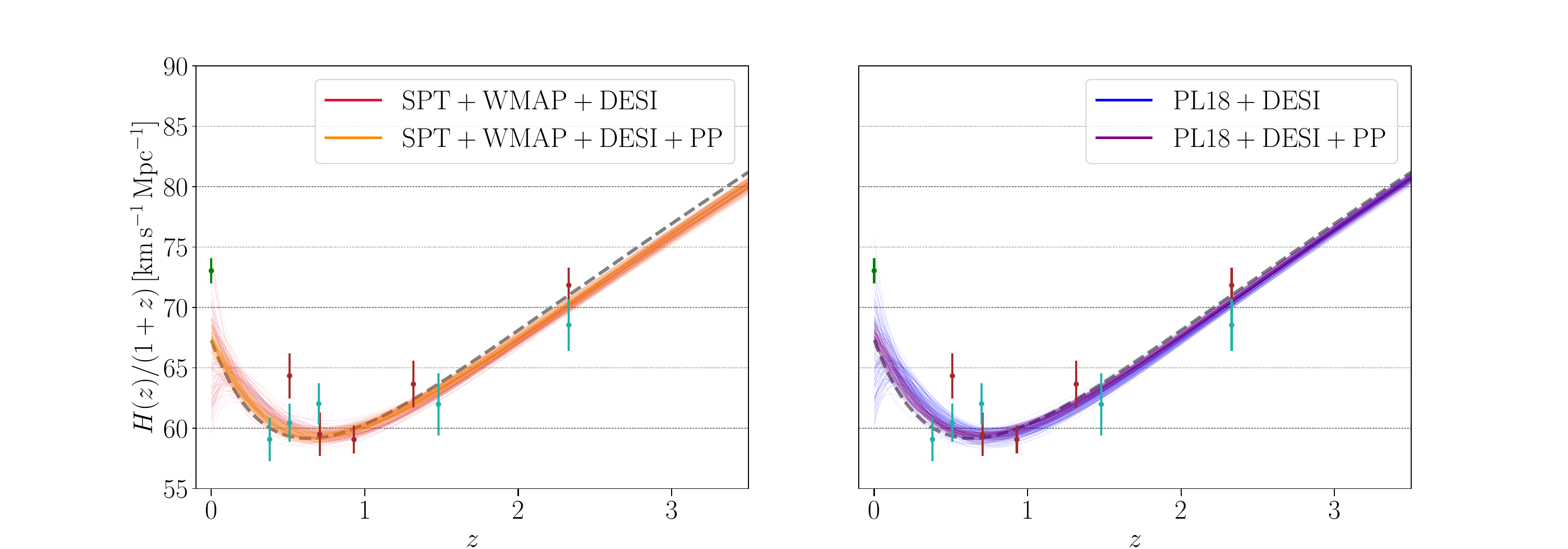}
     \caption{ODE constraints (top panels post-filtered results, bottom panels full-prior results) on the redshift evolution of the $H(z)/(1+z)$ function, derived from the \texttt{SPT+WMAP+DESI} and \texttt{PL18+DESI} data combinations, with and without supernovae. Also shown are $D_H/r_d$ measurements from \texttt{eBOSS}~\cite{eBOSS:2020yzd} (green) and \texttt{DESI YR1}~\cite{DESI:2024mwx} (red), obtained using luminous red galaxies, emission line galaxies, and quasars as tracers; the point at $z=0$ is the SH0ES constraint $H_0=73.04\pm1.04 \,\rm km\,s^{-1}\,Mpc^{-1}$~\cite{Riess:2021jrx}. These measurements are converted into $H(z)$ using the Planck constraint $r_d = 147.09 \pm 0.26\,\, {\rm Mpc}$. The grey dashed line represents the $\Lambda$CDM best-fit curve from \texttt{TTTEEE+lowE+lensing}~\cite{Planck:2018vyg}.
     }
    \label{fig:h}
\end{figure*}


\begin{figure*}[t]
\makebox[\textwidth][c]{
    \includegraphics[width=1.1\textwidth]{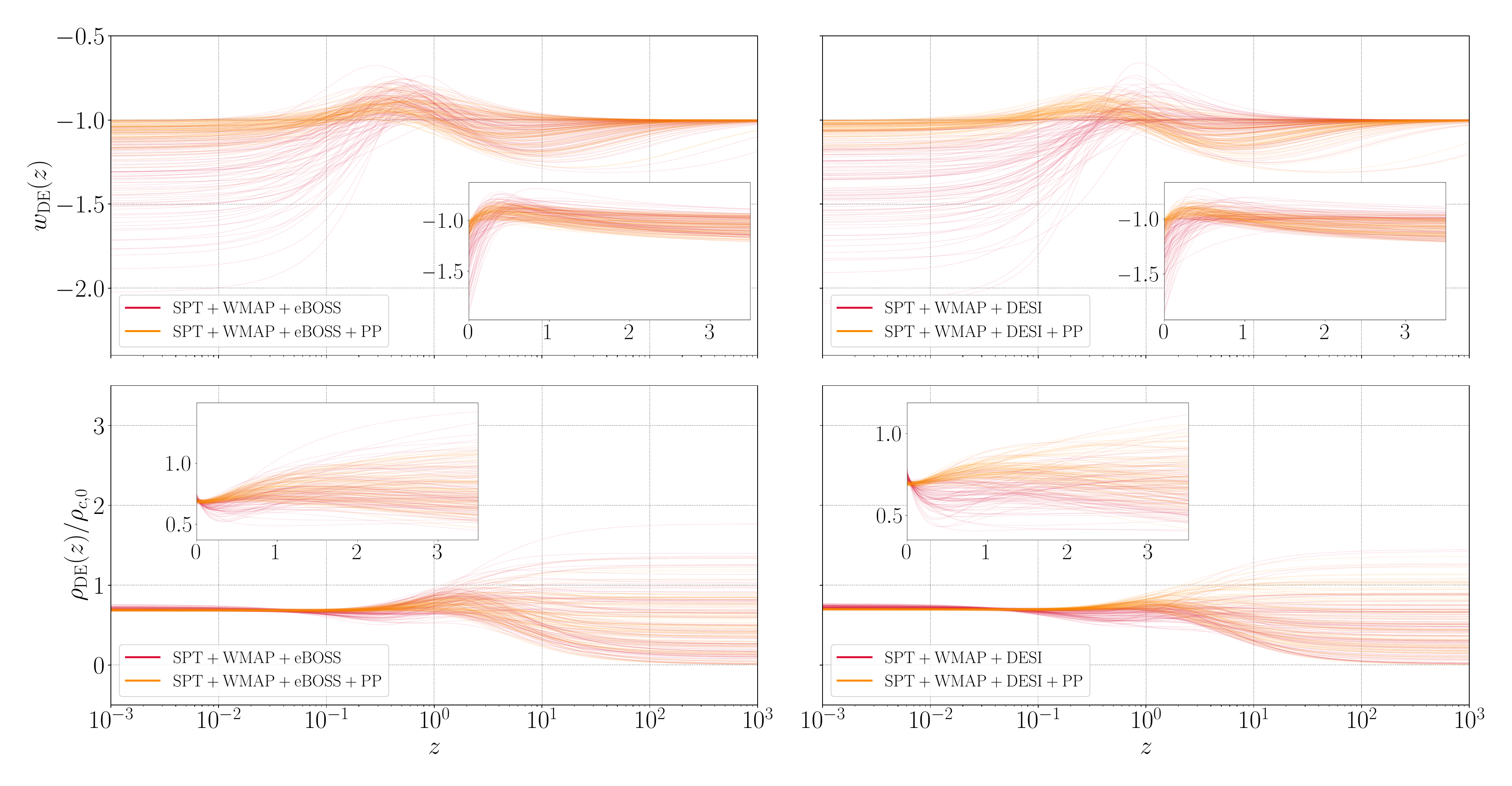}}
     \caption{Redshift evolution of $w_{\rm DE}(z)$ (top panels) and $\rho_{\rm DE}(z)/\rho_{c,0}$ (bottom panels) obtained from the post-filtered results using \texttt{SPT} as the baseline CMB dataset. In each panel the reader can find the same plot contained in the panel itself, re-expressed through a linear but shorter redshift range.}
    \label{fig:wrho_spt}

\end{figure*}


\begin{figure*}[t]
\makebox[\textwidth][c]{
    \includegraphics[width=1.1\textwidth]{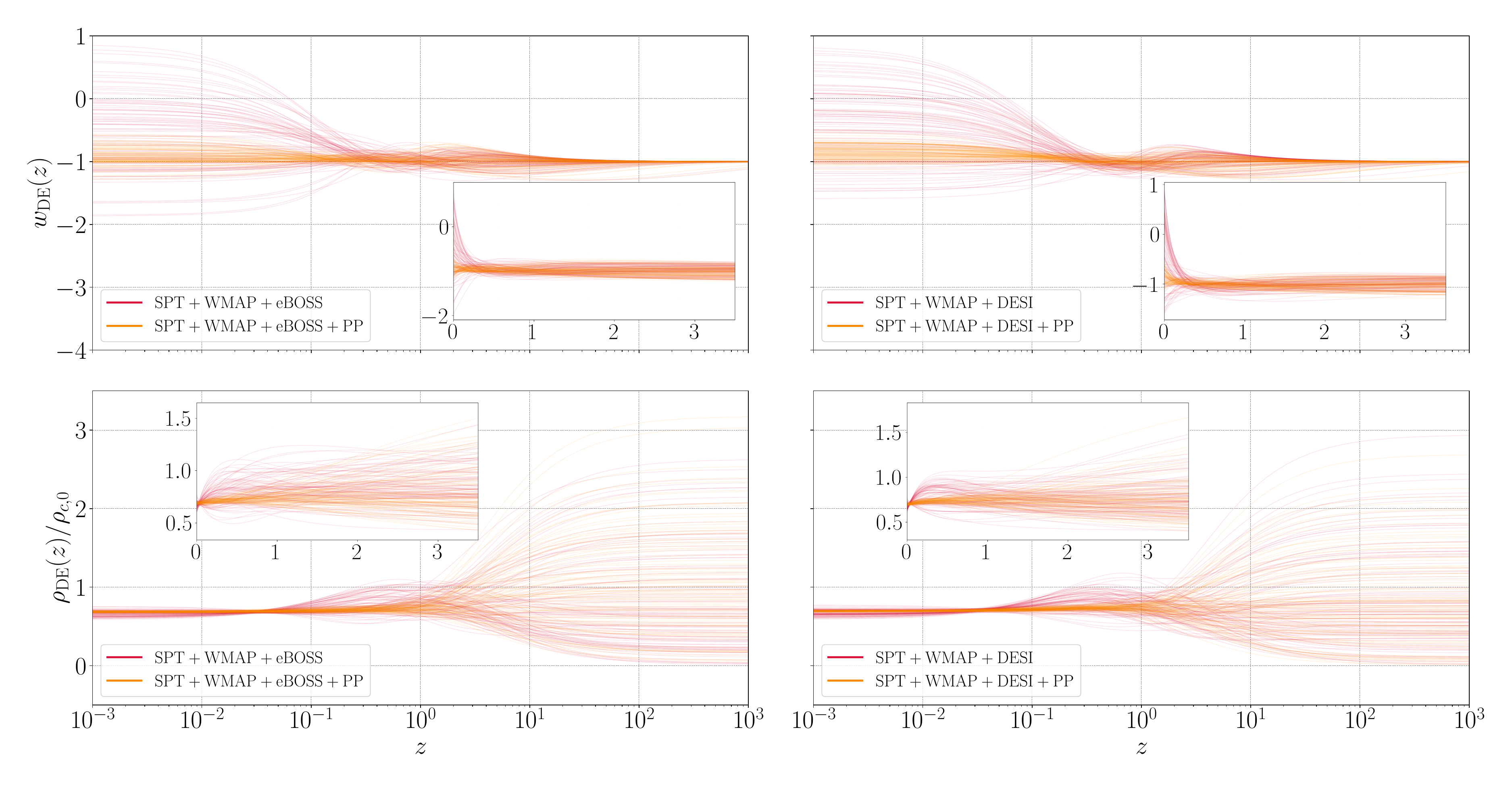}}
     \caption{Same as Fig.~\ref{fig:wrho_spt} but for the full-prior results.}
    \label{fig:wrho_sptfull}

\end{figure*}


\begin{figure*}[t]
\makebox[\textwidth][c]{    \includegraphics[width=1.1\textwidth]{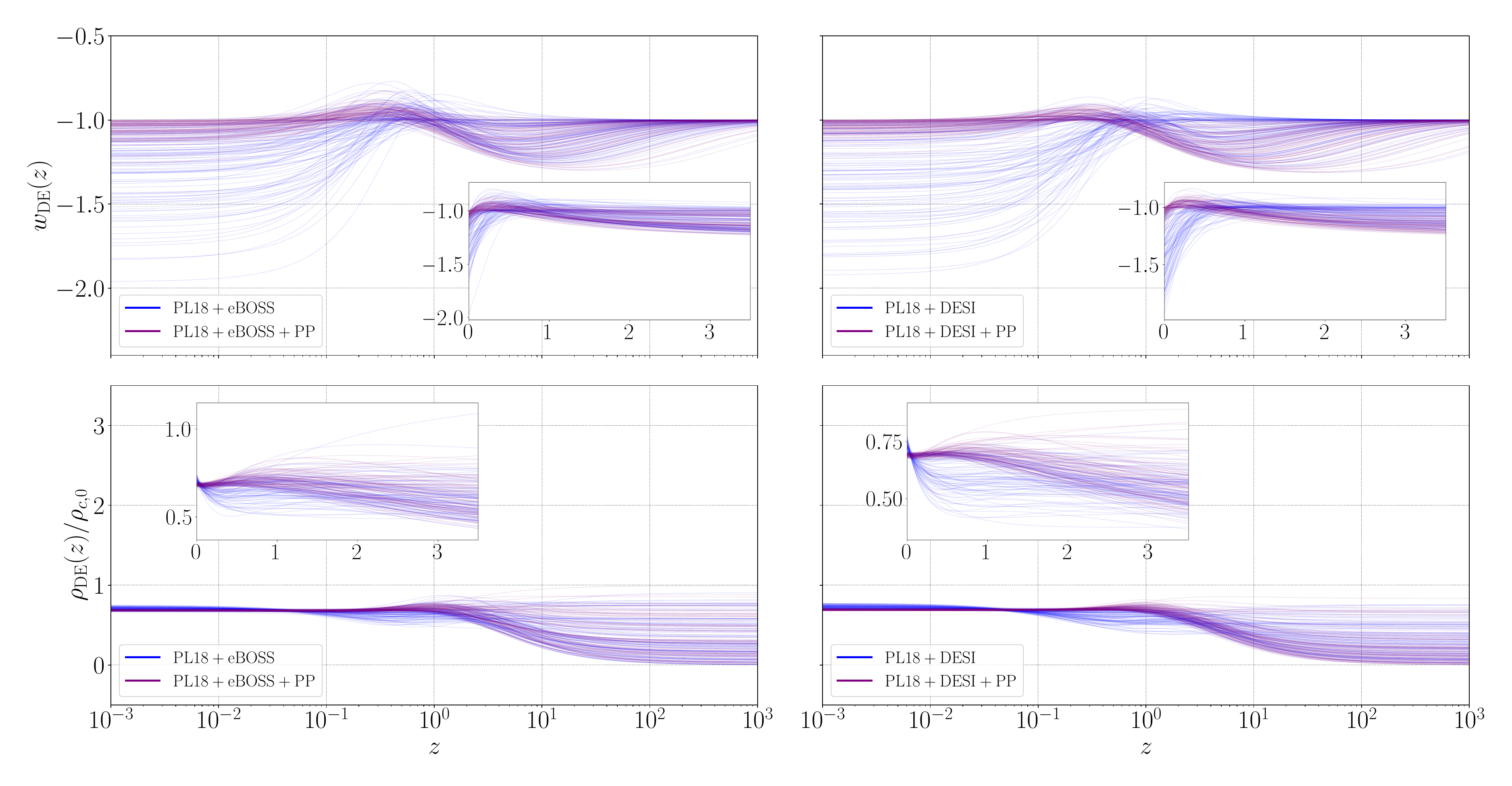}}
     \caption{Redshift evolution of $w_{\rm DE}(z)$ (top panels) and $\rho_{\rm DE}(z)/\rho_{c,0}$ (bottom panels) obtained from the post-filtered results using \texttt{PL18} as the baseline CMB dataset. In each panel the reader can find the same plot contained in the panel itself, re-expressed through a linear but shorter redshift range.}
    \label{fig:wrho_pl18}
\end{figure*}


\begin{figure*}[t]
\makebox[\textwidth][c]{    \includegraphics[width=1.1\textwidth]{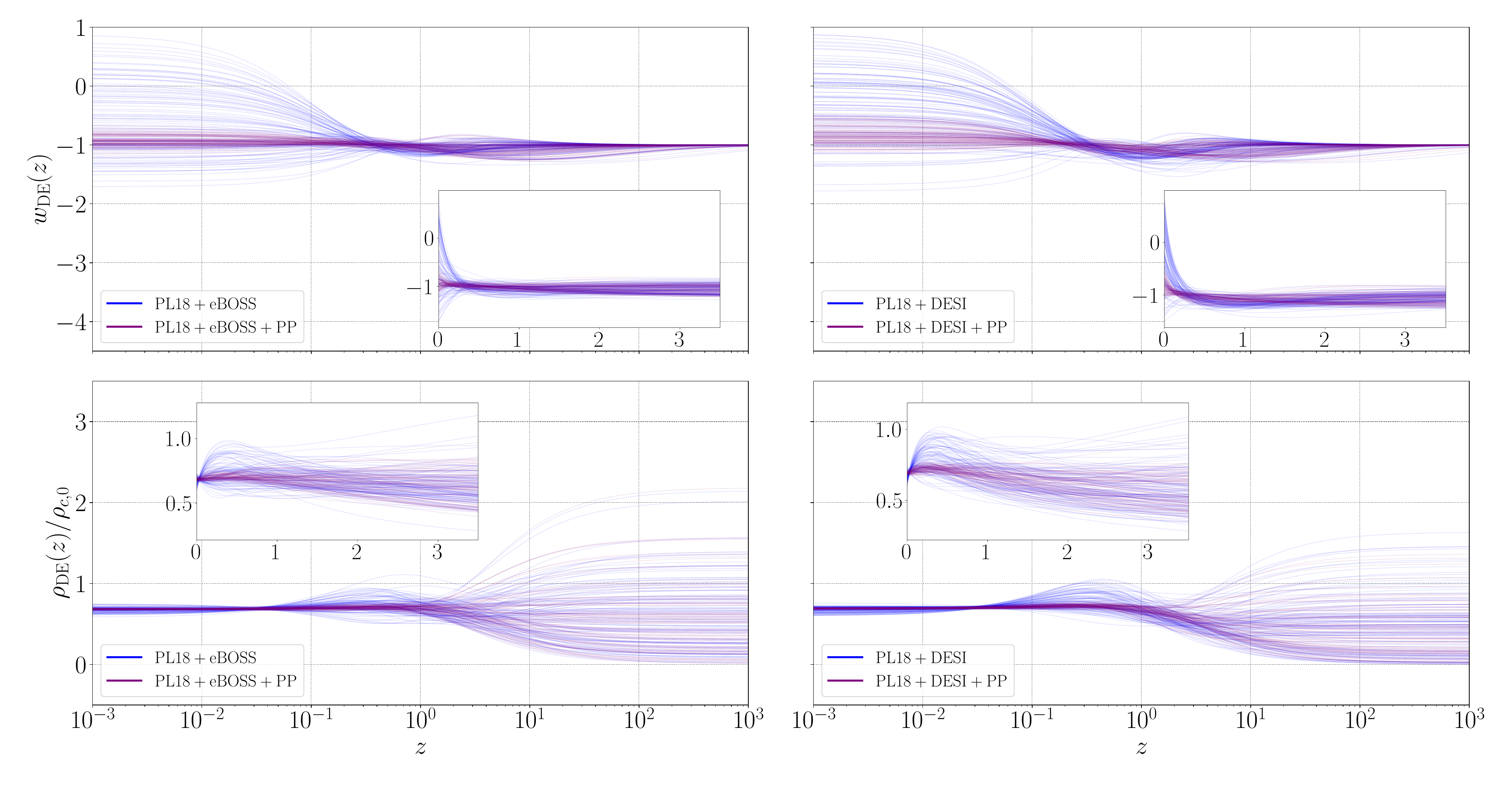}}
     \caption{Same as Fig.~\ref{fig:wrho_pl18} but for the full-prior results.}
    \label{fig:wrho_pl18full}
\end{figure*}


\section{Results and Discussion}
\label{results}

\begin{table*}[t]
\centering
\caption{68\% CL constraints on the free parameters (above the line) and derived parameters (below the line), obtained from the post-filtered results using \texttt{SPT} as the baseline CMB dataset. $BE$ is the Bayesian Evidence.}
\label{tab:spt_cut}
\begin{tabular}{|l|c|c|c|c|}
\hline
 & SPT+WMAP & SPT+WMAP & SPT+WMAP & SPT+WMAP  \\
 & +eBOSS & +eBOSS+PP & +DESI & +DESI+PP  \\
\hline

\boldmath$\Omega_b h^2$ &  $ 0.02242\pm 0.00021 $&$ 0.02241\pm 0.00020 $&$ 0.02244\pm 0.00020 $&$ 0.02244\pm 0.00019$
\\

\boldmath$\Omega_c h^2$ &  $ 0.1163^{+0.0021}_{-0.0019} $&$ 0.1167\pm 0.0019 $&$ 0.1157^{+0.0016}_{-0.0014} $&$ 0.1157\pm 0.0016$
\\

\boldmath$100\theta_{MC} $ &  $ 1.04018\pm 0.00065 $&$ 1.04015\pm 0.00063 $&$ 1.04028\pm 0.00062 $&$ 1.04029\pm 0.00064$
\\

\boldmath$\tau$ &  $ 0.0532\pm 0.0072 $&$ 0.0532\pm 0.0072 $&$ 0.0538\pm 0.0070 $&$ 0.0537\pm 0.0070$
\\

\boldmath${\rm{ln}}(10^{10} A_s)$ &  $ 3.031\pm 0.015 $&$ 3.032\pm 0.015 $&$ 3.030\pm 0.015 $&$ 3.030\pm 0.015$
\\

\boldmath$n_s$ &  $ 0.9679\pm 0.0061 $&$ 0.9673\pm 0.0057 $&$ 0.9690\pm 0.0055 $&$ 0.9691\pm 0.0055$
\\

$\alpha$ &  $ < 3.12 $&$ 1.54^{+0.66}_{-0.94} $&$ < 1.80 $&$ 1.40^{+0.65}_{-0.84}$
\\

$\beta$ &  $ < 3.94 $&$ 1.88^{+0.81}_{-1.5} $&$ < 2.98 $&$ 1.88^{+0.95}_{-1.2}$
\\

$a_m$ &  $ 0.826^{+0.13}_{-0.069} $&$ 0.934^{+0.056}_{-0.023} $&$ 0.74^{+0.16}_{-0.12} $&$ > 0.930$
\\

\hline

$\Omega_m$ &  $ 0.296^{+0.022}_{-0.016} $&$ 0.3113\pm 0.0070 $&$ 0.276^{+0.025}_{-0.016} $&$ 0.3030\pm 0.0069$
\\

$H_0$ [km/s/Mpc] &  $ 68.8^{+1.7}_{-2.7} $&$ 67.02\pm 0.68 $&$ 71.0^{+1.8}_{-3.2} $&$ 67.68\pm 0.71$
\\

$S_8$ &  $ 0.787^{+0.024}_{-0.021} $&$ 0.798\pm 0.020 $&$ 0.775^{+0.020}_{-0.017} $&$ 0.786\pm 0.018$
\\

$r_{d}$ [Mpc] &  $ 148.03\pm 0.53 $&$ 147.93\pm 0.50 $&$ 148.16\pm 0.44 $&$ 148.18\pm 0.45$
\\

\hline

$BE$ & $-0.31$ & $1.04$ & $-0.95$ & $0.84$
\\
\hline
\end{tabular}
\end{table*}

\begin{table*}[t]
\centering
\caption{68\% CL constraints on the free parameters (above the line) and derived parameters (below the line), obtained from the full-prior results using \texttt{SPT} as the baseline CMB dataset. $BE$ is the Bayesian Evidence.}
\label{tab:sptfull}

\begin{tabular}{|l|c|c|c|c|}
\hline
 & SPT+WMAP & SPT+WMAP & SPT+WMAP & SPT+WMAP  \\
 & +eBOSS & +eBOSS+PP & +DESI & +DESI+PP   \\
\hline

\boldmath$\Omega_b h^2$ &  $ 0.02242\pm 0.00021 $&$ 0.02241\pm 0.00020 $&$ 0.02243\pm 0.00020 $&$ 0.02244\pm 0.00020$
\\

\boldmath$\Omega_c h^2$ &  $ 0.1167\pm 0.0020 $&$ 0.1169^{+0.0020}_{-0.0017} $&$ 0.1159\pm 0.0017 $&$ 0.1158\pm 0.0016$
\\

\boldmath$100\theta_{MC} $ &  $ 1.04014\pm 0.00065 $&$ 1.04012\pm 0.00065 $&$ 1.04026\pm 0.00064 $&$ 1.04029\pm 0.00064$
\\

\boldmath$\tau$ &  $ 0.0531\pm 0.0071 $&$ 0.0531\pm 0.0072 $&$ 0.0535\pm 0.0072 $&$ 0.0536\pm 0.0072$
\\

\boldmath${\rm{ln}}(10^{10} A_s)$ &  $ 3.032\pm 0.015 $&$ 3.032\pm 0.015 $&$ 3.030\pm 0.015 $&$ 3.030\pm 0.015$
\\

\boldmath$n_s$ &  $ 0.9674\pm 0.0059 $&$ 0.9670\pm 0.0058 $&$ 0.9686\pm 0.0056 $&$ 0.9689\pm 0.0055$
\\

$\alpha$ &  $ 0.6^{+1.9}_{-2.6} $&$ 0.2^{+1.0}_{-1.2} $&$ 0.6^{+2.3}_{-3.1} $&$ 0.1^{+1.0}_{-1.3}$
\\

$\beta$ &  $ -0.7^{+3.9}_{-3.5} $&$ -0.3^{+2.5}_{-1.5} $&$ -2.6^{+2.3}_{-4.6} $&$ -0.7^{+2.5}_{-1.5}$
\\

$a_m$ &  $ 0.66^{+0.26}_{-0.32} $&$ 0.79\pm 0.27 $&$ 0.58\pm 0.28 $&$ 0.73\pm 0.28$
\\

\hline

$\Omega_m$ &  $ 0.330^{+0.029}_{-0.038} $&$ 0.3129\pm 0.0073 $&$ 0.323^{+0.036}_{-0.041} $&$ 0.3048\pm 0.0071$
\\

$H_0$ [km/s/Mpc] &  $ 65.4\pm 3.3 $&$ 66.90\pm 0.68 $&$ 65.9^{+3.6}_{-4.2} $&$ 67.50\pm 0.70$
\\

$S_8$ &  $ 0.804\pm 0.025 $&$ 0.801^{+0.022}_{-0.019} $&$ 0.796\pm 0.024 $&$ 0.788\pm 0.018$
\\

$r_{d}$ [Mpc] &  $ 147.94\pm 0.53 $&$ 147.88\pm 0.51 $&$ 148.12\pm 0.47 $&$ 148.16\pm 0.46$
\\

\hline

$BE$ & $0.43$ & $0.67$ & $0.25$ & $1.62$
\\
\hline

\end{tabular}
\end{table*}

\begin{table*}[t]
\centering
\caption{68\% CL constraints on the free parameters (above the line) and derived parameters (below the line), obtained from the post-filtered results using \texttt{PL18} as the baseline CMB dataset. $BE$ is the Bayesian Evidence.}
\label{tab:pl18_cut}

\resizebox{0.75\textwidth}{!}{%
\begin{tabular}{|l|c|c|c|c|}
\hline
 & PL18+eBOSS & PL18+eBOSS+PP & PL18+DESI & PL18+DESI+PP \\
\hline

\boldmath$\Omega_b h^2$ &  $ 0.02242\pm 0.00013 $&$ 0.02241\pm 0.00014 $&$ 0.02248\pm 0.00014 $&$ 0.02247\pm 0.00014$
\\

\boldmath$\Omega_c h^2$ &  $ 0.1195\pm 0.0010 $&$ 0.1195\pm 0.0010 $&$ 0.11860\pm 0.00093 $&$ 0.11883\pm 0.00098$
\\

\boldmath$100\theta_{MC} $ &  $ 1.04099\pm 0.00030 $&$ 1.04097\pm 0.00031 $&$ 1.04113\pm 0.00028 $&$ 1.04107^{+0.00030}_{-0.00026}$
\\

\boldmath$\tau$ &  $ 0.0553\pm 0.0072 $&$ 0.0554\pm 0.0072 $&$ 0.0574\pm 0.0076 $&$ 0.0571\pm 0.0077$
\\

\boldmath${\rm{ln}}(10^{10} A_s)$ &  $ 3.046\pm 0.014 $&$ 3.046\pm 0.014 $&$ 3.048\pm 0.015 $&$ 3.048\pm 0.015$
\\

\boldmath$n_s$ &  $ 0.9664\pm 0.0038 $&$ 0.9665\pm 0.0038 $&$ 0.9688\pm 0.0038 $&$ 0.9683\pm 0.0038$
\\

$\alpha$ &  $ < 1.50 $&$ < 1.25 $&$ < 1.02 $&$ < 1.01$
\\

$\beta$ &  $ 2.44^{+0.91}_{-2.1} $&$ 1.71^{+0.84}_{-1.0} $&$ < 2.84 $&$ 1.66\pm 0.72$
\\

$a_m$ &  $ 0.76^{+0.17}_{-0.11} $&$ > 0.904 $&$ 0.66^{+0.26}_{-0.13} $&$ > 0.913$
\\
\hline
$\Omega_m$ &  $ 0.297^{+0.020}_{-0.013} $&$ 0.3146\pm 0.0066 $&$ 0.277^{+0.030}_{-0.019} $&$ 0.3071^{+0.0058}_{-0.0067}$
\\

$H_0$ [km/s/Mpc] &  $ 69.4^{+1.4}_{-2.4} $&$ 67.33^{+0.69}_{-0.62} $&$ 71.7^{+2.3}_{-3.9} $&$ 67.99^{+0.68}_{-0.56}$
\\

$S_8$ &  $ 0.821\pm 0.012 $&$ 0.828\pm 0.010 $&$ 0.806^{+0.015}_{-0.013} $&$ 0.821\pm 0.010$
\\

$r_{d}$ [Mpc] &  $ 147.17\pm 0.24 $&$ 147.19\pm 0.24 $&$ 147.35\pm 0.23 $&$ 147.31\pm 0.24$
\\
\hline

$BE$ & $7.02$ & $7.59$ & $7.55$ & $9.28$
\\
\hline

\end{tabular}%
}
\end{table*}


\begin{table*}[t]
\centering
\caption{68\% CL constraints on the free parameters (above the line) and derived parameters (below the line), obtained from the full-prior results using \texttt{PL18} as the baseline CMB dataset. $BE$ is the Bayesian Evidence.}
\label{tab:pl18full}
\resizebox{0.75\textwidth}{!}{%
\begin{tabular}{|l|c|c|c|c|}
\hline
 & PL18+eBOSS & PL18+eBOSS+PP & PL18+DESI & PL18+DESI+PP \\
\hline

\boldmath$\Omega_b h^2$ &  $ 0.02241\pm 0.00014 $&$ 0.02241\pm 0.00014 $&$ 0.02245\pm 0.00014 $&$ 0.02246\pm 0.00014$
\\

\boldmath$\Omega_c h^2$ &  $ 0.1196\pm 0.0010 $&$ 0.11959\pm 0.00099 $&$ 0.11904\pm 0.00098 $&$ 0.11889\pm 0.00097$
\\

\boldmath$100\theta_{MC} $ &  $ 1.04096\pm 0.00030 $&$ 1.04097\pm 0.00030 $&$ 1.04105\pm 0.00029 $&$ 1.04107\pm 0.00029$
\\

\boldmath$\tau$ &  $ 0.0549\pm 0.0073 $&$ 0.0551\pm 0.0073 $&$ 0.0560\pm 0.0075 $&$ 0.0566\pm 0.0075$
\\

\boldmath${\rm{ln}}(10^{10} A_s)$ &  $ 3.045\pm 0.014 $&$ 3.045\pm 0.014 $&$ 3.046\pm 0.015 $&$ 3.047\pm 0.015$
\\

\boldmath$n_s$ &  $ 0.9662\pm 0.0038 $&$ 0.9663\pm 0.0038 $&$ 0.9677\pm 0.0038 $&$ 0.9681\pm 0.0038$
\\

$\alpha$ &  $ 0.6^{+1.9}_{-3.3} $&$ 0.02^{+0.91}_{-1.2} $&$ 1.1\pm 3.9 $&$ 0.2^{+1.2}_{-1.9}$
\\

$\beta$ &  $ -1.5^{+3.9}_{-3.2} $&$ 0.1^{+1.8}_{-1.1} $&$ < -2.12 $&$ -0.5^{+2.3}_{-1.0}$
\\

$a_m$ &  $ 0.58^{+0.28}_{-0.54} $&$ 0.75^{+0.28}_{-0.24} $&$ < 0.743 $&$ < 0.824$
\\

\hline

$\Omega_m$ &  $ 0.334^{+0.030}_{-0.039} $&$ 0.3165\pm 0.0068 $&$ 0.329\pm 0.036 $&$ 0.3093\pm 0.0067$
\\

$H_0$ [km/s/Mpc] &  $ 65.6\pm 3.3 $&$ 67.14\pm 0.66 $&$ 66.1^{+3.1}_{-4.2} $&$ 67.77\pm 0.69$
\\

$S_8$ &  $ 0.836\pm 0.016 $&$ 0.830\pm 0.010 $&$ 0.831^{+0.018}_{-0.016} $&$ 0.823\pm 0.010$
\\

$r_{d}$ [Mpc] &  $ 147.15\pm 0.24 $&$ 147.17\pm 0.24 $&$ 147.27\pm 0.23 $&$ 147.29\pm 0.23$
\\

\hline

$BE$ & $7.65$ & $7.90$ & $9.84$ & $9.59$
\\
\hline
\end{tabular}%
}
\end{table*}

We begin our discussion by justifying our choice of priors for the extra parameters of DMS20, namely, $\{a_m, \alpha, \beta\}$. In contrast to Refs.~\cite{Adil:2023exv,DiValentino:2020naf}, the prior ranges for $\{\alpha, \beta\}$ are chosen to include negative values, and the prior range for $a_m$ is extended to include values greater than $1$. In Refs.~\cite{Adil:2023exv,DiValentino:2020naf}, the choice of priors enforces $\alpha > 0$, $\beta > 0$, and $a_m \in [0,1]$. This restricts $\rho_{\rm DE}$ to an evolution history with certain properties, the most important being the guaranteed existence of a PDL crossing at $a = a_m$, where the EoS parameter transitions from quintessence ($w_{\rm DE} > -1$) to phantom ($w_{\rm DE} < -1$) as the universe expands. However, in both of the previous studies, the reported posteriors for these parameters do not exclude the prior bounds for many of the dataset combinations, particularly those that do not include the SH0ES $H_0$ measurement. Moreover, recent indications of a dynamical dark energy component suggest evidence for a preference of a PDL crossing in the opposite direction,\footnote{It is important to note that the prior ranges chosen in Refs.~\cite{Adil:2023exv,DiValentino:2020naf} allow (but do not guarantee) for a second PDL crossing in this opposite direction in addition to the one at $a = a_m$, and even a third crossing may occur in a discontinuous way if the DE density attains negative values in the past. In fact, when the model is constrained with data, all three of these crossings were found to play an important role in the evolution of the DE density.} i.e., where the EoS parameter transitions from phantom ($w_{\rm DE} < -1$) to quintessence ($w_{\rm DE} > -1$) as the universe expands~\cite{DESI:2024mwx,DESI:2025zgx}. In fact, the PDL crossing at $a = a_m$ occurs in this opposite direction when $\alpha < 0$. More generally, various qualitatively different DE density evolution histories become available in the DMS20 model when the parameters are allowed to explore this extended parameter space; see Table II and Fig. 1 of Ref.~\cite{Adil:2023exv} and the discussions therein.

In this work, we chose the extended priors in~\cref{tab:priors} to take a more conservative approach that does not guarantee the existence of $a_m$ within the expansion history, and allowed the MCMC analysis to explore the parameter space of the model more freely. Indeed, the results we report suggest that these extended regions of parameter space are always consistent with our dataset combinations, and are sometimes even preferred. To better compare our results with previous works, we also report results obtained after filtering the MCMC chains by dropping any samples with $\alpha < 0$, $\beta < 0$, or $a_m > 1$. We ensured that all of the filtered chains satisfied the Gelman–Rubin convergence criterion with $R - 1 < 0.07$; this approach is equivalent to performing the analysis with the corresponding restricted priors. We refer to the results from the full chain as the ``full-prior results,'' and to those from the post-processed chains as the ``post-filtered results.''

In this section, we present the main results of our analysis using different combinations of datasets. We analyze constraints on our dynamical dark energy model ODE, characterized by the parameters $\alpha$, $\beta$, and $a_m$, using various combinations of the CMB and BAO measurements described in the previous section, with and without supernovae distance moduli. 

In~\cref{fig:cut,fig:full}, we present the parameter constraints derived from both the full-prior and post-filtered analyses. We find that using full priors results in multimodal posterior distributions for certain parameters, while the post-filtered analysis yields unimodal distributions. The multimodality in the full results arises from the phenomenological degeneracies in the parameter space, i.e., different regions of the parameter space correspond to similar expansion histories; see Tab.II in~\cite{Adil:2023exv}. The $\Lambda$CDM model is nested in DMS20 and corresponds to the $\alpha=\beta=0$ section of the parameter space. It is seen from~\cref{fig:full} that there is always consistency, in all of the data set combinations shown in the triangular plot, with the part of the parameter space corresponding to $\Lambda$CDM, and there is no evidence for dynamics in the DE density evolution. Overall, despite the different prior choices (to be discussed in detail later), the results remain statistically consistent across all analyses. A comparative analysis of the \texttt{SPT+WMAP+DESI+PP} and \texttt{PL18+DESI+PP} combinations shows that the latter yields tighter parameter constraints, as illustrated by the red and blue contour plots. Moreover, the inferred value of $S_8$ is systematically lower when using the \texttt{SPT+WMAP+DESI+PP} combination compared to \texttt{PL18+DESI+PP}. Substituting \texttt{DESI} with \texttt{eBOSS} leads to additional shifts in the best-fit values of several parameters (see the following discussions for the respective $\sigma$ deviations), highlighting the sensitivity of the results to the choice of BAO dataset.
 
We present the complete set of parameter estimates from our analysis in~\cref{tab:spt_cut,tab:pl18_cut,tab:sptfull,tab:pl18full}. Post-filtered results show significantly tighter constraints on cosmological parameters compared to the full-prior setup. In both the \texttt{SPT} and \texttt{PL18} cases, the parameters $\alpha$, $\beta$, and $a_m$ are more precisely determined when the prior ranges on model parameters are restricted. This is expected as the post-filtering removes a significant portion that is within 68$\%$ CL of the mean values of the full-prior marginalized posteriors. The inclusion of \texttt{PP} 
data further enhances the precision of these constraints. For instance, the \texttt{SPT+WMAP+DESI+PP} combination with post-filtered priors (see Table~\ref{tab:spt_cut}) yields, at 68\% CL, $\alpha = 1.40^{+0.65}_{-0.84}$, $\beta = 1.88^{+0.95}_{-1.2}$, and $a_m > 0.930$. In contrast, the same dataset under the full-prior setup gives $\alpha = 0.1^{+1.0}_{-1.3}$, $\beta = -0.7^{+2.5}_{-1.5}$, and $a_m = 0.73 \pm 0.28$ at 68\% CL (see Table~\ref{tab:sptfull}). This corresponds to deviations of $1.06\sigma$, $0.93\sigma$, and $0.71\sigma$ for $\alpha$, $\beta$, and $a_m$, respectively, when comparing the prior cut-off case with the full-prior analysis for the same dataset combinations.

While the full-prior results are better for interpreting the true phenomena that our data sets prefer, the post-filtered results are easier to interpret and match the previous studies~\cite{DiValentino:2020naf,Adil:2023exv}. The post-filtered results enforce the existence of a DE EoS parameter that crosses from a quintessence regime to phantom regime at $a=a_m$. For instance, we see in~\cref{tab:pl18_cut} that \texttt{PL18+eBOSS} yields a clear peak for the scale of the PDL crossing at $a_m = 0.76^{+0.17}_{-0.11}$ and the addition of \texttt{PP} yields $a_m > 0.904 $. This is accompanied by a reduction in the uncertainty of $H_0$, which changes from $H_0=69.4^{+1.4}_{-2.4} \,\,\rm km/s/Mpc$ to $H_0=67.33^{+0.69}_{-0.62} \,\,\rm km/s/Mpc$ with $0.83\sigma$ deviations, closer to Planck's $\Lambda$CDM value.

Interestingly, replacing the eBOSS data with the newer DESI dataset further tightens the constraints on the ODE parameters ($\alpha$, $\beta$) and $H_0$. The Hubble constant shifts slightly toward higher values, reflecting the preference of DESI for a lower matter density relative to eBOSS, and the well-known anti-correlation between $H_0$ and $\Omega_m$. We note that a PDL crossing from $w_{\rm DE}<-1$ to $w_{\rm DE}>-1$ always seems to occur in our post-filtered results, and it is then followed by a second PDL crossing at $a=a_m$ (see \cref{fig:wrho_spt,fig:wrho_pl18}), which is well constrained in most cases.

We observe that for the \texttt{SPT+WMAP+eBOSS} combination, the mean value of $H_0$ is noticeably lower compared to the corresponding result with \texttt{PL18}. However, replacing eBOSS with DESI leads to a significant shift of the Hubble constant towards higher values, with $H_0 = 71.0^{+1.8}_{-3.2}$ km/s/Mpc, in agreement with local measurements with $1.56\sigma$ deviations as compared to $\sim5\sigma$ for the $\Lambda$CDM case, along with a correspondingly lower matter density. Moreover, as in the results with \texttt{PL18}, the addition of \texttt{PP} improves the constraints on $\alpha,\beta$, slightly deviating from $\alpha,\beta=0$ ($\Lambda$CDM). It also increases the preference for the ODE model over $\Lambda$CDM, in both the filtered and the full-prior cases, as one can see from the bayesian evidence values in Tables \ref{tab:spt_cut}, \ref{tab:sptfull}, \ref{tab:pl18_cut} and \ref{tab:pl18full}. However, such preference is not substantial when \texttt{SPT} is used instead of \texttt{Planck} as the CMB benchmark dataset, with the former case showing only slightly positive bayesian evidence values when \texttt{SPT} CMB data are used alongside the Pantheon+ supernovae measurements.

Different behaviors are observed when analyzing the model with the full priors. As shown in Tables \ref{tab:sptfull} and~\ref{tab:pl18full}, $\beta$ consistently prefers a mean value in the negative region, with the 68$\%$ CL for the \texttt{SPT+WMAP+DESI} case being negative, and the \texttt{PL18+DESI} case having an upper bound of $-2.12$. Hence, we see that the post-filtering forces the $\beta$ values to be larger. One could expect, in parallel, that post-filtering would shift the $a_m$ constraints to lower values by dropping the $a_m>1$ samples in the chain. However, in contrast, the favored values of $a_m$ are smaller for the full-prior, resulting in an upper limit when DESI data are used with Planck. This is due to the correlations between $\{\alpha,\beta,a_m\}$. Finally, although $\alpha$ appears to have significantly larger uncertainties, Fig.~\ref{fig:full} shows that its posterior is bimodal around zero for the any dataset combination including \texttt{PP} (except for the \texttt{SPT+WMAP+eBOSS+PP} case), with the peak for the negative values corresponding to a larger probability.

In our analysis, the inferred values of the Hubble constant $H_0$ remain consistently lower than the SH0ES measurement ($H_0 = 73.04 \pm 1.04$ km/s/Mpc), though the extent of the discrepancy varies across datasets. For example, the \texttt{SPT+WMAP+eBOSS} combination yields $H_0=68.8^{+1.7}_{-2.7}$ km/s/Mpc with post-filtered priors (Table~\ref{tab:spt_cut}), and $H_0 = 65.4\pm 3.3$ km/s/Mpc with full priors (Table~\ref{tab:sptfull}) having the deviations of $2.13\sigma$ and $2.21\sigma$ respectively with the inferred SH0ES value. In contrast, the \texttt{PL18+DESI+PP} combination provides a more precise full-prior estimate of $H_0 = 67.77 \pm 0.69$ km/s/Mpc (Table~\ref{tab:pl18full}) with the deviation of $4.12\sigma$ with SH0ES value and $0.43\sigma$ with Planck value. 
Although certain configurations formally yield a significantly reduced tension with the SH0ES measurement (e.g., the analyses without \texttt{PP}), these tension reductions are mainly due to increased uncertainties of the $H_0$ constraints, a direct consequence of the three extra parameters of the DMS20 model. Nevertheless, it is worth noting that both the \texttt{SPT+WMAP+DESI} and \texttt{PL18+DESI} post-filtered results reduce the discrepancy with SH0ES to roughly $1\sigma$, due to both a genuine shift of the mean $H_0$ value toward higher values and the increased uncertainties.

Top panels of~\cref{fig:h} illustrate the evolution of the Hubble parameter for the \texttt{PL18+DESI}, \texttt{PL18+DESI+PP}, \texttt{SPT+WMAP+DESI}, and \texttt{SPT+WMAP+DESI+PP} data set combinations of the post-filtered results, while the bottom panels present the corresponding results for the full-prior case. While all these scenarios, along with the baseline $\Lambda$CDM model, remain discrepant with the DESI LRG measurement at redshift $z=0.510$, we can see some improvement in the full-prior results. These figures show, as it was also discussed in the previous paragraph, that
the post-filtered results for \texttt{PL18+DESI} and \texttt{SPT+WMAP+DESI} reduce the $H_0$ tension relative to $\Lambda$CDM albeit with large uncertainties. As shown in the figures, the inclusion of \texttt{PP} data reduces the uncertainty and lowers the mean value of $H_0$, putting it in strong tension with the SH0ES measurement. Comparing the top panels with the bottom ones, it is evident that restricting the priors to satisfy $\alpha>0$ and $\beta>0$ significantly limits the flexibility of DMS20 to describe different expansion histories.

Next, we present the evolution of the dark energy density, $\rho_{\rm DE}(z)$, and the corresponding EoS parameter, $w_{\rm DE}(z)$, derived from the posterior chains of our analysis, as shown in~\cref{fig:wrho_spt,fig:wrho_pl18} for the post-filtered results, and in~\cref{fig:wrho_sptfull,fig:wrho_pl18full} for the full-prior ones. While noticeable deviations from a cosmological constant ($w = -1$) are permitted across all results, such behavior is not necessarily favored, as discussed previously. 
A further striking feature shared by all results is the generic occurrence of a double PDL crossing, characterized by opposite crossing directions in each instance. However, the full-prior and post-filtered results differ noticeably, as the order of the PDL crossings may be reversed between them. Specifically, the post-filtered results show an initial crossing into the quintessence regime, followed by a second crossing into the phantom regime. This ordering is by construction, as the post-filtered results assume the existence of a PDL crossing from the quintessence to the phantom regime, which must be the latest crossing if multiple ones occur. In contrast, the full-prior results may instead exhibit an initial crossing into the phantom regime, followed by a second one into the quintessence regime—although they can also display the same crossing pattern as the post-filtered results. However, note that, while the double PDL crossing is generic, it is not present in all the MCMC samples for either full or post-filtered results. A double crossing for the post-filtered results would correspond to a initially increasing DE density that goes through a decreasing phase before falling back to a decreasing regime; while this behavior describes the majority of the samples in~\cref{fig:wrho_spt,fig:wrho_pl18}, DE density evolutions that start with an initially increasing phase are also ubiquitous, especially when \texttt{DESI} is not present in the data set. Similar observations can be made for the full-prior results. As a corollary to these discussions, we see that the present day value of the EoS parameter is always in the phantom regime for the post-filtered results by construction, but it is mostly in the quintessence regime for the full-prior results especially when \texttt{DESI} is present in the data set combination. Addition of \texttt{PP} exacerbates this behavior further. This late-time preference for a quintessence EoS parameter obtained from combining CMB, \texttt{DESI} and \texttt{PP} data agrees with the findings of~\cite{DESI:2024mwx,DESI:2025zgx}. To quantify this preference (it also serves as a measure to quantify deviations from $\Lambda$CDM), we report constraints on the present-day dark energy equation of state parameter $w_0$ in Table~\ref{tab:w0_values} for the full-prior results. The inferred values range from $w_0 = -0.93^{+0.11}_{-0.13}$ for the \texttt{SPT+WMAP+eBOSS+PP} combination to $w_0 = -0.34^{+0.53}_{-0.73}$ for \texttt{PL18+DESI}, indicating, depending on the specific case, a mild preference for quintessence-like behavior within the $2\sigma$ range. Although $w_0 = -1$ remains consistent with all datasets at the 95\% CL, the central values point to slight deviations from a pure cosmological constant, in agreement with the evolving trends in $\rho_{\rm DE}(z)$ and $w_{\rm DE}(z)$ discussed earlier.

\begin{table}[h]
    \centering
    \caption{Values of $w_0$ for different dataset combinations (full-prior results). Uncertainties correspond to 68\% CL, with 95\% CL shown in parentheses.}
    \renewcommand{\arraystretch}{1.2}
    \resizebox{0.85\columnwidth}{!}{%
    \begin{tabular}{|l|c|}
        \hline
        \texttt{Dataset} & \texttt{$w_0$} \\
        \hline
        PL18+eBOSS & $-0.55^{+0.50\ (+1.3)}_{-0.71\ (-0.99)}$ \\
        PL18+eBOSS+PP & $-0.92^{+0.087\ (+0.25)}_{-0.12\ (-0.20)}$ \\
        PL18+DESI & $-0.34^{+0.53\ (+1.2)}_{-0.73\ (-1.0)}$ \\
        PL18+DESI+PP & $-0.848^{+0.097\ (+0.27)}_{-0.16\ (-0.22)}$ \\
        SPT+WMAP+eBOSS & $-0.65^{+0.44\ (+1.3)}_{-0.71\ (-1.0)}$ \\
        SPT+WMAP+eBOSS+PP & $-0.93^{+0.11\ (+0.27)}_{-0.13\ (-0.23)}$ \\
        SPT+WMAP+DESI & $-0.44^{+0.58\ (+1.3)}_{-0.75\ (-1.0)}$ \\
        SPT+WMAP+DESI+PP & $-0.86^{+0.11\ (+0.27)}_{-0.15\ (-0.22)}$ \\
        \hline
    \end{tabular}
    }
    \label{tab:w0_values}
\end{table}

\section{Conclusion}
\label{final}

In this work, we have carried out a comprehensive analysis of a dynamical dark energy model parameterized by the parameters $\{\alpha, \beta, a_m\}$, extending their prior ranges beyond those considered in previous studies, to enable a broader exploration of the parameter space of the model. This extended framework allows for qualitatively richer evolution histories of the dark energy density $\rho_{\rm DE}(z)$ and equation of state $w_{\rm DE}(z)$, including possible phantom divide line (PDL) crossings in both directions. 
By combining a variety of observational datasets—including CMB measurements from SPT, WMAP, and Planck; BAO data from eBOSS and DESI; and supernovae observations from Pantheon$+$ (PP)—we find mild deviations from a cosmological constant at intermediate redshifts, as our results are generally consistent with $\Lambda$CDM within 2$\sigma$ or less. In particular, the EoS generically exhibits two PDL crossings in opposite directions within a single expansion history. These features are robust across different dataset combinations, suggesting that dynamical dark energy with physically nontrivial features might be present in a concordance model.

We also investigated the impact of prior choices by comparing the constraints obtained from full-prior and post-filtered MCMC chains. The post-filtered results assume the presence of a PDL crossing from quintessence to phantom regime, and are better comparable to Refs.~\cite{DiValentino:2020naf,Adil:2023exv} that analyze the same DE model. The full priors reveal multimodal behavior highlighting the critical role of prior selection in capturing the full phenomenology of the model, as in this case the preferred present-day value of the EoS parameter is in the quintessence regime unlike the post-filtered results, which assume a present-day phantom value. Both cases are statistically consistent with -1.
For some data set combinations the $H_0$ tension is ameliorated formally due to significantly enlarged uncertainties in comparison to the $\Lambda$CDM model, however the combinations of our CMB, BAO and supernovae data give tight constraints on $H_0$, in strong tension with the SH0ES measurements.

A robust qualitative feature that emerges across essentially all of our likelihood combinations is the frequent appearance of multiple (regular) crossings of the phantom divide line (PDL), $w_{\rm DE}=-1$, within a single dark-energy evolution history. If this phenomenology persists with future data, it would carry nontrivial implications for fundamental model building, but it is crucial to state precisely what is excluded. In particular, a minimally coupled single scalar field with a standard kinetic structure, ${\cal L}_\phi=\epsilon X - V(\phi)$ (with $\epsilon=+1$ for quintessence and $\epsilon=-1$ for a pure phantom), cannot realize a \emph{stable} crossing of $w_{\rm DE}=-1$ while maintaining $\rho_{\rm DE}>0$: such models satisfy $w_{\rm DE}\ge -1$ (quintessence) or $w_{\rm DE}\le -1$ (phantom) as long as $\rho_{\rm DE}>0$ (see, e.g., Ref.~\cite{Akarsu:2025gwi}). Therefore, even a single smooth PDL crossing with strictly positive $\rho_{\rm DE}$ already points beyond the vanilla quintessence/phantom paradigm. Multiple crossings---especially when accompanied by oscillatory behavior and, in parts of parameter space, sign changes in $\rho_{\rm DE}$---tighten these requirements further and naturally motivate multi-degree-of-freedom dark-sector realizations (e.g.\ \emph{quintom} or effective multi-fluid descriptions) and/or an effective origin from modified gravity\footnote{We note, however, that if $\rho_{\rm DE}$ is allowed to cross zero, a single-field phantom setup can accommodate transitions from $\rho_{\rm DE}<0$ with $w_{\rm DE}>-1$ to $\rho_{\rm DE}>0$ with $w_{\rm DE}< -1$ at late times; see Ref.~\cite{Akarsu:2025gwi}. In this case the divergence of $w_{\rm DE}=p_{\rm DE}/\rho_{\rm DE}$ at $\rho_{\rm DE}=0$ is a kinematic artifact of the ratio, rather than a singularity of the background expansion; see, e.g., Refs.~\cite{Akarsu:2026anp,Akarsu:2025gwi}.}. Importantly, this discussion should \emph{not} be interpreted as excluding general single-field constructions: more elaborate frameworks---for instance non-minimally coupled scalars or single-field EFTs with non-trivial kinetic structure/braiding and broken shift symmetry---can yield an \emph{effective} equation of state that crosses $-1$ and may fit current data well (see, e.g., Refs.~\cite{Wolf:2025jed,Wolf:2025acj}). Our point is instead that the specific combination of features suggested by our constraints makes the simplest minimally coupled canonical realizations inadequate and provides concrete motivation to pursue fundamental ODE-like realizations, including non-canonical fields (e.g.\ \cite{Franchino-Vinas:2019nqy,Franchino-Vinas:2021bcl}), coupled systems, or emergent dark-sector phenomena. Additionally, as a complementary diagnostic, one may map the ODE phenomenology onto an effective CPL description in the $(w_0,w_a)$ plane by treating $(w_0,w_a)$ as a data-compression of the model predictions: for a range of ODE parameter choices, one can generate the corresponding observables and determine the best-fitting $(w_0,w_a)$ while verifying that the residuals remain small (e.g.\ \cite{Garcia-Garcia:2019cvr}). For a meaningful comparison with data-driven constraints in the $(w_0,w_a)$ plane, the mapping should also account for the redshift sensitivity and error budget of the probes employed (e.g.\ \cite{Wolf:2023uno,Wolf:2024eph}). A dedicated implementation of this programme is beyond the scope of the present analysis and is left to future work.

\section{Acknowledgments}
SAA acknowledges the support of the DGAPA postdoctoral fellowship program at ICF-UNAM, Mexico.
EDV is supported by a Royal Society Dorothy Hodgkin Research Fellowship. RCN thanks the financial support from the Conselho Nacional de Desenvolvimento Científico e Tecnologico (CNPq, National Council for Scientific and Technological Development) under the project No. 304306/2022-3, and the Fundação de Amparo à Pesquisa do Estado do RS (FAPERGS, Research Support Foundation of the State of RS) for partial financial support under the project No. 23/2551-0000848-3. \"{O}.A. acknowledges the support by the Turkish Academy of Sciences in the scheme of the Outstanding Young Scientist Award  (T\"{U}BA-GEB\.{I}P). AAS acknowledges the funding from ANRF, Govt. of India, under the research grant no. CRG/2023/003984. This article is based upon work from COST Action CA21136 Addressing observational tensions in cosmology with systematics and fundamental physics (CosmoVerse) supported by COST (European Cooperation in Science and Technology). The authors acknowledge the use of High-Performance Computing resources from the IT Services at the University of Sheffield.

\bibliography{bibliorev}
\end{document}